\documentclass[sigconf,10pt]{acmart}

\usepackage{adjustbox}
\usepackage[table]{xcolor}
\usepackage{siunitx}
\usepackage{booktabs}

\makeatletter
\AtBeginDocument{\let\balance\relax}
\makeatother

\AtBeginDocument{%
  }

\setcopyright{acmlicensed}
\copyrightyear{2026}
\acmYear{2026}
\acmDOI{}

\acmYear{2026}\copyrightyear{2026}
\acmConference[EuroMLSys '26]{The 6th Workshop on Machine Learning and Systems}{April 27--30, 2026}{Edinburgh, Scotland Uk}
\acmBooktitle{The 6th Workshop on Machine Learning and Systems (EuroMLSys '26), April 27--30, 2026, Edinburgh, Scotland Uk}
\acmDOI{10.1145/3805621.3807636}
\acmISBN{979-8-4007-2605-7/26/04}

\begin{document}

\title{Systems-Level Attack Surface of Edge Agent Deployments on IoT}

%% Authors
\author{Zhonghao Zhan}
\affiliation{%
  \institution{Imperial College London}
  \city{London}
  \country{UK}
}

\author{Krinos Li}
\affiliation{%
  \institution{Imperial College London}
  \city{London}
  \country{UK}
}

\author{Yefan Zhang}
\affiliation{%
  \institution{ByteDance}
  \city{Seattle}
  \state{WA}
  \country{USA}
}

\author{Hamed Haddadi}
\affiliation{%
  \institution{Imperial College London}
  \city{London}
  \country{UK}
}

%%
%% Abstract
%%
\begin{abstract}
Edge deployment of LLM agents on IoT hardware introduces attack surfaces absent from cloud-hosted orchestration. We present an empirical security analysis of three architectures (cloud-hosted, edge-local swarm, and hybrid) using a multi-device home-automation testbed with local MQTT messaging and an Android smartphone as an edge inference node. We identify five systems-level attack surfaces, including two emergent failures observed during live testbed operation: coordination-state divergence and induced trust erosion. We frame core security properties as measurable systems metrics: data egress volume, failover window exposure, sovereignty boundary integrity, and provenance chain completeness. Our measurements show that edge-local deployments eliminate routine cloud data exposure but silently degrade sovereignty when fallback mechanisms trigger, with boundary crossings invisible at the application layer. Provenance chains remain complete under cooperative operation yet are trivially bypassed without cryptographic enforcement. Failover windows create transient blind spots exploitable for unauthorised actuation. These results demonstrate that deployment architecture, not just model or prompt design, is a primary determinant of security risk in agent-controlled IoT systems.
\end{abstract}

%%
%% CCS Concepts — generate at https://dl.acm.org/ccs/ccs.cfm
%%


%%
%% Keywords
%%

\settopmatter{printacmref=true, authorsperrow=3}

\maketitle

%%
%% Body
%%
\section{Introduction}

LLM-based agents are rapidly moving from cloud APIs to edge hardware controlling physical devices. Commercial deployments now include SwitchBot AI Hub (a smart-home hub advertising local AI capabilities)~\cite{switchbot2026aihub}, Home Assistant's MCP integration exposing tools to conversation agents~\cite{homeassistantMCP}, and distributed frameworks like OpenClaw running on consumer hardware~\cite{openclaw}. This shift changes the threat model fundamentally: where cloud-hosted agents present a single point of compromise with inherent data exfiltration, edge deployments distribute both capability and attack surface across heterogeneous nodes~\cite{roman2018}.

Recent deployments and incidents highlight the urgency. OpenClaw's official security guidance explicitly centers prompt injection, gateway exposure, and trusted-operator assumptions as first-order risks in personal-assistant deployments~\cite{openclaw2025}. The 7,000-unit DJI Romo vacuum breach (Feb 2026) demonstrated how compromised credentials grant mass control over physically distributed agents~\cite{dji2026}. Yet the systems community has not systematically examined how deployment architecture shapes agent security for IoT.

Critically, these agents do not choose their communication
infrastructure: they \emph{inherit} it.  MQTT pub/sub is a widely used coordination
substrate in IoT deployments~\cite{mishra2020, andy2017, harsha2018}.  When LLM agents are deployed on
this hardware, MQTT becomes their command-and-control plane by
default.  Our analysis examines the security consequences of this
inheritance: what happens when a protocol designed for telemetry
becomes the coordination bus for autonomous agents with physical
actuation.

Existing agent security research focuses on prompt injection~\cite{greshake2023, debenedetti2024}, tool-use risk evaluation~\cite{ruan2024}, and agent reasoning or self-improvement scaffolds~\cite{shinn2023, yao2023react}, typically in centralized digital environments rather than edge-local IoT deployments~\cite{zhan2024}. These defenses do not transfer cleanly to edge deployments where: (i)~agents run on resource-constrained hardware unable to execute full guardrail stacks locally; (ii)~network partitions isolate edge nodes, creating audit gaps; (iii)~local data sovereignty eliminates cloud exfiltration risk but enables lateral movement between co-located agents; and (iv)~physical actuation latency becomes a security-relevant metric.

We ask: \textbf{how does deployment architecture change the systems-level attack surface of agent-controlled IoT?} Through empirical measurements on our testbed, we make three contributions:

\begin{enumerate}
    \item \textbf{Architectural threat characterization.} We define three deployment patterns as distinct threat models: cloud-hosted (single point, inherent egress), edge-local swarm (distributed surface, zero egress), and hybrid (inherited risks from both), and enumerate five attack surfaces, including coordination-state divergence and induced trust erosion, two emergent failures observed during testbed operation.

    \item \textbf{Systems metrics as security properties.} We measure actuation-to-audit delay, provenance chain completeness, data sovereignty (external egress), and failover vulnerability windows, demonstrating that these systems-level quantities determine the feasibility of real-time safety monitoring, forensic analysis, and attack interception.

    \item \textbf{Empirical findings.} The full end-to-end actuation pipeline (MQTT publish through Home Assistant actuation to confirmation) achieves ${\sim}117$\,ms mean delay (P95 ${<}152$\,ms), enables complete provenance for cooperative agents but provides zero cryptographic enforcement, eliminates external data egress entirely during normal operation but silently degrades when fallback mechanisms trigger, and exposes vulnerability windows during network failover.
\end{enumerate}

\begin{figure}[t]
  \centering
  \includegraphics[width=\linewidth]{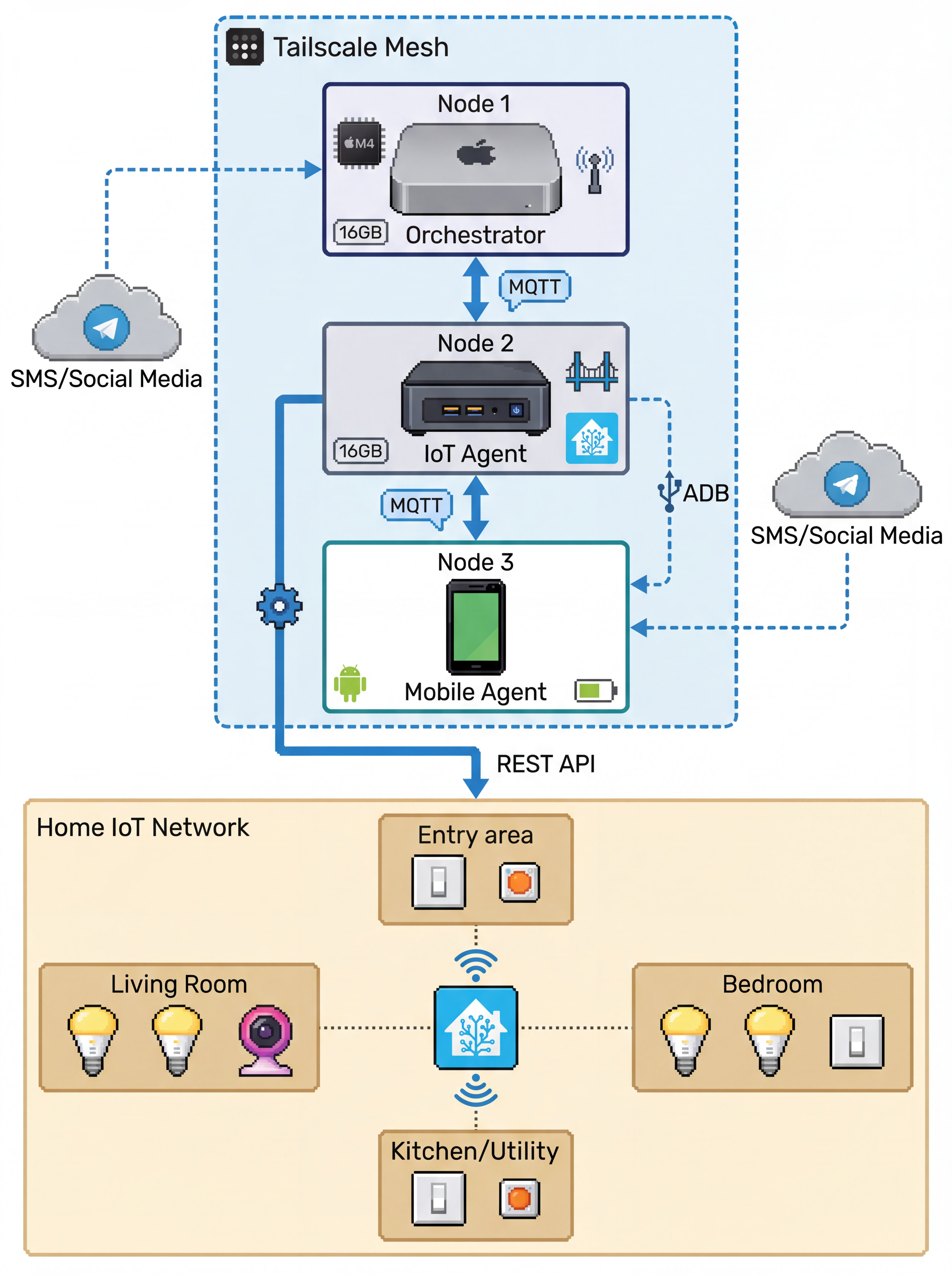}
  \caption{Testbed topology.  Inter-agent traffic traverses the
    Tailscale mesh via MQTT pub/sub on the Mac mini.  The NUC
    bridges MQTT to Home Assistant for IoT actuation.  WAN links
    carry only LLM inference and Telegram traffic.}
  \Description{Network topology diagram showing three agent nodes
    connected via Tailscale VPN mesh with MQTT messaging, plus
    external API connections for inference and messaging.}
  \label{fig:topology}
\end{figure}
\section{Threat Model and Architecture}

\subsection{Deployment Architectures}

We identify three deployment patterns for LLM agents on IoT, each
producing a distinct security posture.

\paragraph{Cloud-Hosted Orchestration.}
The agent runtime (inference, tool calls, decision loop) executes in
cloud infrastructure; IoT devices are controlled via cloud-to-device
APIs.  All sensor data, commands, and reasoning context transit
externally.  The architecture provides centralized audit and a single
enforcement point, but necessarily places operational data on external infrastructure and expands confidentiality and trust dependencies~\cite{roman2018}.

\paragraph{Edge-Local Swarm.}
Multiple agents run on local hardware, coordinating via a local
message broker with no cloud dependency for normal operation.
Operational data never leaves the local network.  The architecture
distributes the attack surface across heterogeneous nodes, requires
per-node compromise, but lacks centralized enforcement and places
resource constraints on local guardrails.

\paragraph{Hybrid (Edge + Cloud).}
Edge agents handle low-latency actuation; cloud handles heavy inference. Unless carefully partitioned, this
pattern inherits the attack surfaces of both: cloud egress for
inference traffic plus a distributed local surface for actuation.

\subsection{Testbed}

We deploy an actively-used edge-local agent swarm with different LLM models utilizing OpenClaw~\cite{openclaw} to control real
IoT hardware (operational daily for home automation and research
coordination):

\begin{itemize}
  \item \textbf{Mac mini M4} (16\,GB, macOS): orchestrator agent
    ``Rupert'' (Claude Opus 4.6), MQTT broker
  \item \textbf{Moto G35} (Android 15, Termux): mobile edge agent
    ``Percy'' (Gemini 3.1 Pro)
  \item \textbf{Intel NUC N150} (16\,GB, Ubuntu 24.04): Home
    Assistant bridge and IoT gateway ``Jeeves'' (GPT-5.2)
  \item \textbf{IoT devices:} Philips Hue lights (5), smart
    switches (3), camera (1), motion sensors (2), speaker (1), microphone(1)
\end{itemize}

\paragraph{MQTT as coordination backbone.}
We do not advocate MQTT as an agent communication protocol; rather,
our testbed inherits MQTT from the IoT ecosystem it controls.
All inter-agent communication uses MQTT pub/sub on the Mac mini
broker (port 1883, Tailscale~\cite{donenfeld2017} mesh only; no public exposure).  Agents
publish to topic-structured channels using a JSON envelope carrying sender ID,
message type, microsecond timestamp, correlation ID, and payload.
The NUC bridges MQTT to Home Assistant's REST API for IoT device
control.  Model inference calls traverse WAN to cloud providers;
all operational IoT traffic remains mesh-local.

This design makes MQTT the \emph{sole coordination plane} for the
swarm: every agent decision, actuation command, status update, and
audit record passes through the broker.  Consequently, the security
properties of the MQTT layer directly determine the security posture
of the entire deployment.

\subsection{Attack Surfaces}
\label{sec:surfaces}

We identify five systems-level attack surfaces that are
\emph{architectural consequences} of edge-local deployment, not
present (or present differently) in cloud-hosted orchestration.  We
exclude model-level attacks (prompt injection, jailbreaking) and supply chain
attacks. These are orthogonal to deployment architecture and
addressed by complementary defenses~\cite{greshake2023}.  We assume three attacker capabilities:
(1)~remote network access (e.g., compromised credentials,
exposed MQTT port); (2)~physical access to an edge node (e.g., stolen phone, USB debugging); (3)~a rogue agent within the
swarm (e.g., compromised skill package installing a malicious MQTT
client).

Table~\ref{tab:surfaces} summarises the five surfaces, their
architectural root causes, and the systems metrics that expose them.

\begin{table}[t]
\centering
\caption{Attack surfaces of edge-local agent deployments.
  Each surface is an architectural consequence of the deployment
  pattern, measurable through systems-level artifacts.}
\label{tab:surfaces}
\begingroup
\setlength{\tabcolsep}{3pt}
\renewcommand{\arraystretch}{1.15}
\rowcolors{2}{gray!08}{white}
\begin{tabular}{p{0.9cm}p{2.0cm}p{1.8cm}p{1.8cm}r}
\toprule
\rowcolor{white}
 Surface & Root Cause & Attack / Effect & Metric & \S \\
\midrule
S1a & No cryptographic binding in MQTT envelope &
  Agent impersonation; command replay &
  Spoofed msg acceptance rate & \ref{sec:provenance} \\
S1b & No shared state plane; message-only coordination &
  Silent context drift; persistent corruption &
  Divergent context copies & \ref{sec:provenance} \\
S1c & No tiered trust in MQTT bus &
  Induced channel distrust; operator lockout &
  Out-of-band recovery required & \ref{sec:provenance} \\
S2 & Sovereignty boundary is runtime, not architectural &
  Silent data exfiltration via fallback &
  DNS egress; bytes to cloud & \ref{sec:sovereignty} \\
S3 & Audit asynchronous; failover creates blind spots &
  Unaudited actuation during blackout &
  Reconnect window (s) & \ref{sec:failover} \\
\bottomrule
\end{tabular}
\endgroup
\end{table}

We group S1a--c as a family reflecting their shared root cause
(unauthenticated MQTT coordination) while distinguishing their
distinct exploitation patterns and operational manifestations.
\paragraph{S1a: Provenance forgery.}
Provenance in our edge swarm is self-reported metadata in the MQTT
JSON envelope with no cryptographic binding~\cite{mishra2020,
firdous2017}.  Any client with broker credentials can impersonate
any agent or publish to safety-critical topics.
\paragraph{S1b: Coordination-state divergence.}
Without a shared state plane, each agent assembles shared context
incrementally from MQTT messages, embedding full copies that drift
silently. We observed
two failure modes: (i)~redundant state duplication (file contents
embedded rather than referenced) and (ii)~invisible semantic
drift (independent modifications with no conflict detection). 
\paragraph{S1c: Induced trust erosion.}
An attacker publishes obviously-forged messages on the MQTT bus.
The defending agent, unable to distinguish forged from legitimate
traffic, treats the entire channel as compromised and refuses
subsequent operator commands.  We observed this during testbed
operation: recovery required out-of-band confirmation via human admin through Telegram.
\paragraph{S2: Silent sovereignty degradation.}
Edge-local deployment eliminates external egress by construction.
However, when an agent falls back to cloud inference (resource
exhaustion, model unavailability), operational data silently
transits externally with no notification or audit record.  The
sovereignty boundary degrades under stress precisely when monitoring
is most needed.
\paragraph{S3: Actuation-audit temporal gap.}
Every command has a delay between physical actuation and the audit
record becoming observable.  During network failover, we observe
multi-second windows where agents actuate with no audit trail,
with the network stack as the dominant bottleneck.

\begin{figure*}[t]
\centering
\includegraphics[width=\linewidth]{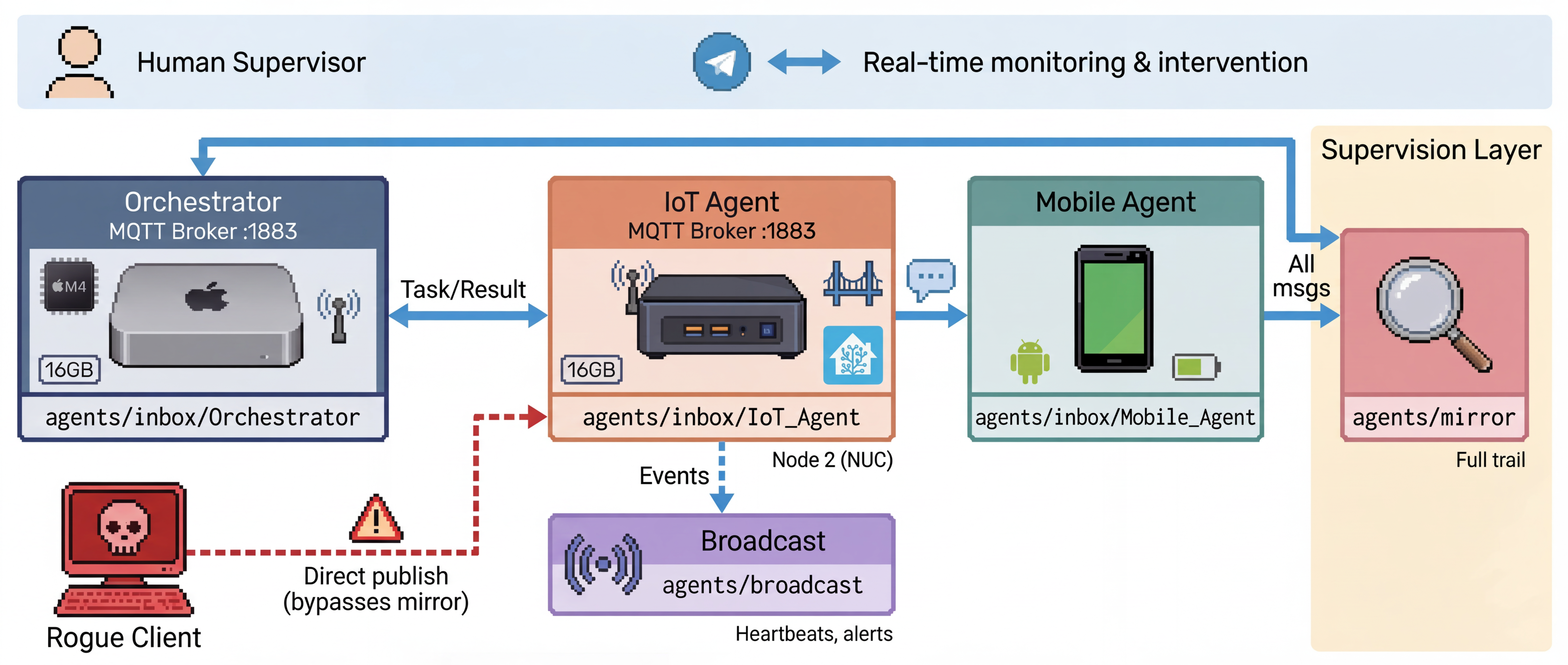}
\caption{MQTT message flow and supervision architecture. Agents communicate via per-agent inbox topics (\texttt{agents/inbox/\{id\}}), with all messages streamed to \texttt{agents/mirror} for real-time human monitoring via Telegram. A rogue client (bottom-left) can publish directly to agent inboxes, bypassing the supervision layer---illustrating the provenance gap measured in Table~\ref{tab:adversarial}.}
\Description{MQTT message flow diagram showing three agent nodes communicating via named inbox topics, a broadcast topic for heartbeats, a supervision layer mirroring all messages to Telegram for human oversight, and a rogue client bypassing the mirror with direct publishes to agent inboxes.}
\label{fig:mqtt-flow}
\end{figure*}

\section{Empirical Analysis}
\label{sec:empirical}

\subsection{Actuation-to-Audit Delay}

For agent-controlled physical systems, unauthorized actions must be detected before physical completion. A door lock command takes ${\sim}500$\,ms to actuate; if the audit trail records the command within that window, a safety monitor can intercept it.

\paragraph{Metric and method.}
We measure the full end-to-end actuation pipeline, not just the MQTT transport hop. An actuation command traverses four stages: (1)~MQTT publish from the commanding agent to the broker, (2)~Home Assistant state read (pre-actuation device check), (3)~Home Assistant service call (physical actuation), and (4)~Home Assistant confirmation (post-actuation state verification). We instrument each stage with microsecond-precision timestamps ($N{=}150$).

\begin{table}[t]
\centering
\caption{End-to-end actuation-to-audit delay: full pipeline from agent command to confirmed device state change ($N{=}150$).}
\label{tab:latency-e2e}
\begingroup
\setlength{\tabcolsep}{3pt}
\renewcommand{\arraystretch}{1.15}
\rowcolors{2}{gray!08}{white}
\begin{tabular}{p{1.45cm}rrrp{1.55cm}}
    \toprule
    \rowcolor{white}
    Stage & Mean & Median & P95 & Note \\
    \midrule
    MQTT publish    & 36.4\,ms  & 36.0\,ms  & 38\,ms  & Agent $\to$ broker \\
    HA state read   & 28.6\,ms  & 29.0\,ms  & 31\,ms  & Pre-actuation check \\
    HA actuation    & 27.6\,ms  & 26.0\,ms  & 64\,ms  & Service call (light toggle) \\
    HA confirmation & 24.1\,ms  & 24.0\,ms  & 31\,ms  & Post-actuation verify \\
    \midrule
    \textbf{Total E2E} & \textbf{116.7\,ms} & \textbf{114.5\,ms} & \textbf{152\,ms} & Full pipeline \\
    \bottomrule
\end{tabular}
\endgroup
\end{table}

\paragraph{Pipeline decomposition.}
The MQTT transport hop accounts for only 31\% of the total delay (Table~\ref{tab:latency-e2e}); the Home Assistant REST bridge and device confirmation contribute the remaining 69\%. The HA actuation stage shows the highest variance (P95 of 64\,ms vs.\ mean of 27.6\,ms), reflecting device-dependent response times. The total E2E delay of ${\sim}117$\,ms is $3.2{\times}$ the MQTT-only figure but remains well within the interception window for electromechanical actuators such as door locks (${\sim}500$\,ms) and motorized valves (${\sim}1$--$2$\,s).

\paragraph{Cross-path validation.}
The NUC $\to$ Mac mini MQTT path exhibits ${\sim}64$\,ms mean latency ($N{=}50$), a $2.7{\times}$ overhead reflecting node heterogeneity (NUC's single-core N150 vs.\ Mac mini's M4). The NUC path shows near-zero variance across payload sizes (64.3--64.5\,ms), suggesting fixed scheduling overhead rather than serialization cost.

\paragraph{Burst resilience.}
Under sustained load on the NUC ($N{=}100$ messages at 128\,B, rapid-fire), mean latency is 49.8\,ms with 0\% degradation, confirming that the MQTT broker handles burst traffic without queuing delays even from the most constrained node.

\paragraph{Implication.}
Real-time safety monitors are feasible for edge-local MQTT deployments. A monitor observes the full actuation pipeline within ${\sim}117$\,ms (P95 ${<}152$\,ms). This is within the actuation window of electromechanical devices such as door locks (${\sim}500$\,ms) and motorized valves (${\sim}1$--$2$\,s), though insufficient for solid-state relays (${<}5$\,ms).

\subsection{Provenance Chain Completeness}
\label{sec:provenance}

Forensic analysis requires full provenance: which agent issued which command, when, and in response to what event.

\paragraph{Cooperative agents.}
We audit 100 agent commands across 5 action types. All messages carry complete metadata: sender ID, ISO-8601 timestamp, UUID correlation ID, message type, and action details. Coverage: 100\% across all fields.

\paragraph{Adversarial testing.}
We probe MQTT's enforcement against malicious clients:

\begin{table}[t]
\centering
\caption{Adversarial MQTT provenance testing: all attacks accepted by the broker.}
\label{tab:adversarial}
\begingroup
\setlength{\tabcolsep}{3pt}
\renewcommand{\arraystretch}{1.15}
\rowcolors{2}{gray!08}{white}
\begin{tabular}{p{2.2cm}p{2.0cm}p{2.0cm}}
    \toprule
    \rowcolor{white}
    Attack & Broker Response & Impact \\
    \midrule
    Missing sender field & Accepted & Untraceable command \\
    Spoofed sender & Accepted & Rogue agent frames others \\
    Replayed message & Accepted & Command re-executed \\
    Direct safety publish & Accepted & Agent logic bypassed \\
    \bottomrule
\end{tabular}
\endgroup
\end{table}

MQTT provides transport-level delivery but \textbf{no application-level security enforcement} (Table~\ref{tab:adversarial}). These gaps are well-documented in telemetry contexts~\cite{firdous2017, harsha2018}; the critical difference here is that MQTT now carries autonomous agent commands to physical actuators. A spoofed ``unlock door'' command is not a corrupted sensor reading: it triggers irreversible physical action within ${\sim}117$\,ms (Table~\ref{tab:latency-e2e}), faster than any human-in-the-loop can intervene. Provenance therefore depends entirely on agent cooperation. Message signing (HMAC/JWT), nonce-based replay protection, and topic ACLs are required yet absent from the default deployment. We did not observe these mechanisms in the open-source agent frameworks we used.

\subsection{Data Sovereignty and Egress}
\label{sec:sovereignty}

We capture all network traffic on the NUC during a 10-minute agent task session (50 MQTT publishes, 200 sensor reads, 30 light commands) and compare against equivalent cloud API calls.

\begin{table}[t]
\centering
\caption{Data egress comparison: edge-local vs.\ cloud architecture. The aggregate row reports a 10-minute session; the normalised row controls for operation count.}
\label{tab:egress}
\begingroup
\setlength{\tabcolsep}{4pt}
\renewcommand{\arraystretch}{1.05}
\rowcolors{2}{gray!08}{white}
\begin{tabular}{p{2.95cm}>{\raggedleft\arraybackslash}p{1.05cm}>{\raggedleft\arraybackslash}p{1.55cm}>{\raggedleft\arraybackslash}p{1.05cm}}
    \toprule
    \rowcolor{white}
    Architecture & External IPs & Bytes Sent & Operations \\
    \midrule
    Edge-local MQTT & \textbf{0} & \textbf{0\,B} & 280 \\
    Cloud (OpenAI API) & 3 & 64,981\,B & 10 \\
    \midrule
    \rowcolor{white}
    \multicolumn{4}{l}{\textit{Normalised per actuation (matched operation type):}} \\
    Edge-local MQTT & --- & ${\sim}$805\,B & 1 \\
    Cloud API & --- & ${\sim}$3,244\,B & 1 \\
    \bottomrule
\end{tabular}
\endgroup
\end{table}

Edge-local deployments exhibit zero \emph{external} data egress during normal operation (Table~\ref{tab:egress}). All communication remains within the Tailscale mesh. To control for the differing operation counts in our aggregate measurement, we also report normalised per-actuation egress for matched operation types: cloud API calls transmit $4.0{\times}$ the bytes of equivalent edge-local operations. The edge-local bytes represent intra-mesh MQTT traffic (never leaving the Tailscale network); the cloud bytes traverse the public internet to inference providers. This eliminates structural data exfiltration as an attack vector: an adversary cannot observe agent context by monitoring cloud API traffic but must compromise a node within the mesh.

\paragraph{Forced fallback: sovereignty under stress.}
To validate S2, we triggered a sovereignty boundary crossing by
sending the mobile agent (Percy) a 109\,KB sensor dump exceeding
its local model's context capacity.  The agent framework's fallback
chain silently routed inference to a cloud provider (Anthropic API).
We captured traffic via \texttt{tcpdump} on the Mac mini, filtering
Percy's Tailscale IP.

\begin{table}[t]
\centering
\caption{Sovereignty boundary crossing: baseline vs.\ forced fallback.}
\label{tab:sovereignty}
\begingroup
\setlength{\tabcolsep}{4pt}
\renewcommand{\arraystretch}{1.05}
\rowcolors{2}{gray!08}{white}
\begin{tabular}{p{3.85cm}>{\raggedleft\arraybackslash}p{1.15cm}>{\raggedleft\arraybackslash}p{2.0cm}}
    \toprule
    \rowcolor{white}
    Metric & Baseline & Fallback \\
    \midrule
    Packets captured & 10{,}473 & 14{,}638 \\
    Packets to/from Percy & 190 & 434 \\
    DNS queries to \texttt{api.anthropic.com} & \textbf{0} & \textbf{10} \\
    Resolved IP & --- & 160.79.104.10 \\
    User notification & N/A & \textbf{None} \\
    MQTT-layer anomaly & N/A & \textbf{None} \\
    \bottomrule
\end{tabular}
\endgroup
\end{table}

The sovereignty violation is invisible at the application layer
(Table~\ref{tab:sovereignty}): MQTT message patterns are identical
before and after fallback, and the agent itself reported only a
``brief glitch''. No indication that operational context had
transited to external infrastructure.  Only DNS-level monitoring
(tcpdump) revealed the boundary crossing.  The local model
returned a \texttt{CANCELLED} error (code 499) confirming resource
exhaustion as the trigger.  This validates the S2 threat: an
adversary can force sovereignty degradation by inducing resource
exhaustion, causing the agent
framework to silently exfiltrate context via legitimate fallback
channels.

\subsection{Failover Vulnerability Windows}
\label{sec:failover}

Mobile edge agents experience network transitions during which they are unreachable and unauditable.

\paragraph{Method.}
Percy connects via WiFi to the MQTT broker (Tailscale tunnel). We disable WiFi, forcing failover to an ADB/USB bridge path. Keep-alive pings at 500\,ms intervals measure the blackout window. We decompose failover into two phases: (a)~network stack recovery (WiFi re-association, Tailscale tunnel re-establishment) and (b)~MQTT reconnection (client-side reconnect to broker).

\paragraph{End-to-end failover.}
35.7\,s from WiFi loss to first successful ping via the ADB fallback path---dominated by network stack recovery (33.6\,s) and ADB bridge establishment.

\paragraph{MQTT reconnection in isolation ($N{=}50$).}
To isolate the MQTT layer, we simulate broker disconnection at the application level (firewall block/unblock) across varying block durations:

\begin{table}[t]
\centering
\caption{MQTT reconnection latency in isolation ($N{=}50$).}
\label{tab:mqtt-reconnect}
\begingroup
\setlength{\tabcolsep}{4pt}
\renewcommand{\arraystretch}{1.05}
\rowcolors{2}{gray!08}{white}
\begin{tabular}{lr}
    \toprule
    \rowcolor{white}
    Metric & Value \\
    \midrule
    Mean reconnection time & 9.3\,ms ($\sigma = 1.9$\,ms) \\
    P95 & 14\,ms \\
    P99 & 17\,ms \\
    Block-duration effect & None (8.5--10.1\,ms) \\
    \bottomrule
\end{tabular}
\endgroup
\end{table}

MQTT reconnection is near-instantaneous (${<}20$\,ms, Table~\ref{tab:mqtt-reconnect}). The 35\,s end-to-end failover window is entirely attributable to the network and VPN stack, not the messaging layer. This has design implications: reducing the vulnerability window requires network-level solutions (fast roaming, multi-path transport) rather than application-level optimizations.

\paragraph{Implication.}
During the 35\,s blackout, a mobile agent cannot receive commands, publish audit logs, or be monitored. If the agent executes a safety-critical action immediately before network loss, the action log is delayed by 35+\,s. An attacker who can induce network disruption gains a predictable blind spot.

\section{Discussion}

\subsection{Architecture IS Security Posture}

Our measurements support a central claim: for agent-controlled physical systems, \textbf{deployment architecture determines security posture more than any model-level or prompt-level mitigation.} Consider the contrast:

\begin{itemize}
    \item A cloud-hosted agent with state-of-the-art prompt guardrails still transmits 65\,KB of context per planning cycle to external infrastructure, a structural exfiltration channel that no amount of alignment tuning can close.
    \item An edge-local agent with no guardrails produces zero external egress and enables ${\sim}117$\,ms end-to-end actuation-to-audit delay, sufficient for real-time interception of most physical actuators.
\end{itemize}

This is not an argument against model-level safety. It is an argument that the systems community has underweighted architectural decisions. The choice of where to run the agent, how to connect it to devices, and what transport to use for audit logs has measurable, quantifiable security consequences that dominate the threat surface for IoT deployments. An open question is guardrail placement: edge hardware cannot run full guardrail stacks locally, suggesting a tiered model with fast local checks and deferred cloud verification for high-stakes actions. Similarly, the 35\,s vulnerability window (\S\ref{sec:failover}) implies that safety-critical actions should be deferred when network stability is uncertain. Integrating network health signals into agent planning remains unexplored.

\subsection{Relation to Non-LLM Distributed Systems}

Many of the vulnerabilities we identify are well-studied in robotics, microservices, and industrial SCADA (Supervisory Control and Data Acquisition) systems~\cite{stouffer2011guide}. They include unauthenticated messaging, state divergence, failover blind spots. Authenticated messaging (TLS client certificates, HMAC-signed payloads), shared-state protocols (distributed consensus), and network fast roaming (multi-path TCP) are mature techniques in those domains. We believe these transfer directly to the MQTT and network layers of LLM agent deployments. What does \emph{not} transfer is the trust model: traditional distributed systems assume deterministic, programmed agents with fixed capabilities, whereas LLM agents exhibit non-deterministic behaviour, interpret natural-language instructions, and can be manipulated via prompt injection. The S1c vulnerability (induced trust erosion) is specific to this class of agent: a deterministic system would not ``decide'' to distrust its communication channel based on message content. Mitigations must therefore combine established distributed-systems techniques at the infrastructure layer with LLM-specific defences at the agent layer. Neither alone is sufficient.

\subsection{The Provenance Gap}

Our adversarial MQTT testing (\S\ref{sec:provenance}) reveals a critical gap: no open-source agent framework provides cryptographic provenance enforcement. In a cooperative swarm this is invisible as all messages arrive with correct metadata. But a single compromised node can impersonate any agent, replay commands, and publish directly to safety-critical topics.

This gap is architectural, not incidental. MQTT was designed as a lightweight pub/sub transport, not a secure command-and-control protocol. Closing the gap requires: (i)~per-agent signing keys with broker-side verification; (ii)~nonce-based replay protection; and (iii)~topic-level ACLs enforced at the broker. These are well-understood techniques, but their absence from the agent frameworks used in our deployment is consistent with security guidance that still assumes a trusted operator boundary or otherwise treats agentic security at a higher level than coordination-bus provenance~\cite{openclaw2025, owaspagentic2025}. Scaling beyond our three-agent testbed to $N$ agents will require formal trust policies: capability delegation and least-privilege topic access. None of these are provided by MQTT natively.

\subsection{Emergent Coordination Failures}

Operating the testbed revealed two failure modes not anticipated in
our initial threat model, both arising from the absence of
coordination-layer security primitives. These are gaps identified as open
problems in recent multi-agent LLM surveys~\cite{guo2024,
talebirad2023}.

\paragraph{State-plane divergence (S1b).}
During multi-agent task coordination, agents maintained inconsistent
local views of shared context.  Without a shared filesystem or
state-synchronisation protocol, agents embedded full task
specifications in MQTT messages rather than referencing canonical
files. It creates divergent copies that drifted within minutes.  One
agent participated in an experiment via MQTT while its Telegram-facing
session remained unaware the experiment had occurred.  This is a
direct consequence of message-based coordination: MQTT carries
messages, not state.  We addressed this operationally by layering a
git-backed shared state plane over the MQTT bus, providing
content-addressed versioning, agent-attributed commits, and explicit
conflict detection.

\paragraph{Induced trust erosion (S1c).}
After a series of provenance-testing messages (S1a adversarial
experiments), the orchestrator agent began treating \emph{all} MQTT
traffic as potentially adversarial, refusing legitimate operator
commands received on the same channel.  Resolution required
out-of-band communication via Telegram, a trusted secondary channel.
The attack requires no actual compromise: an adversary need only
publish forged messages to degrade operator trust in the
coordination bus.  This is a denial-of-service via \emph{induced
paranoia}, exploiting the absence of tiered trust in MQTT.

\subsection{Silent Sovereignty Degradation}

Our forced-fallback experiment (\S\ref{sec:empirical},
Table~\ref{tab:sovereignty}) confirms that edge-local data sovereignty
is a runtime property, not an architectural guarantee.  The boundary
crossing was invisible at every layer except DNS: MQTT messages showed
no anomaly, the agent reported only a ``brief glitch,'' and no audit
record indicated that operational context had left the local network. Mitigation requires sovereignty-aware fallback policies:
either blocking cloud fallback entirely for sensitive contexts, or
injecting explicit audit markers when the boundary is crossed.

\subsection{Limitations}

Our measurements derive from a single three-node testbed with one
MQTT broker, one VPN substrate (Tailscale), and one IoT platform
(Home Assistant); generalisation to other topologies, brokers, and
network substrates remains future work.  The cloud egress comparison
(Table~\ref{tab:egress}) is not workload-matched: the cloud baseline
captures only inference API calls, not a full cloud-orchestrated
workload performing identical tasks, and serves as a directional
baseline demonstrating structural egress differences rather than a
controlled experiment.  We do not implement or evaluate mitigation
prototypes (e.g., HMAC signing, topic ACLs, sovereignty-aware
fallback policies); our contribution is identifying and measuring the
attack surfaces, not closing them.  The three S1-family surfaces
(S1a--c) share a common root cause and could be viewed as facets of a single
vulnerability rather than independent attack surfaces; we
distinguish them because their exploitation patterns and operational
manifestations differ significantly.  Our failover measurements reflect a specific WiFi$\to$ADB bridge path, which is primarily an Android developer debugging mechanism rather than a consumer failover scenario. In typical deployments, a mobile agent would fail over from WiFi to a 4G/5G cellular network, likely yielding shorter blackout windows due to faster network re-association. Our measurement therefore represents a conservative upper bound; the structural vulnerability (unaudited actuation during any network transition) holds regardless of the specific failover path.

\section{Related Work}
\paragraph{LLM agent security.}
Greshake et al.~\cite{greshake2023} and Zhan et al.~\cite{zhan2024}
demonstrated indirect prompt injection against tool-integrated LLM
applications and agents. Xi et al.~\cite{xi2023} surveyed LLM-based
agents and outlined their architectural components and open problems,
but did not address edge deployment constraints.
AgentDojo~\cite{debenedetti2024} and LM-emulated
sandboxes~\cite{ruan2024} evaluate agent vulnerabilities in controlled
settings; OWASP~\cite{owaspagentic2025} and PASB~\cite{pasb2025}
catalogue and benchmark agentic risks. These works focus on what agents
do wrong (tool misuse, prompt compliance); we complement them by
examining how the infrastructure agents inherit shapes the attack surface
independent of model behaviour.
\paragraph{Edge AI, IoT, and MQTT security.}
Edge computing and MEC surveys~\cite{shi2016, roman2018} discuss
latency, locality, and security/privacy tradeoffs; we measure one such
tradeoff empirically and show that sovereignty can degrade silently
under resource exhaustion. IoT security work has
focused on device-level vulnerabilities~\cite{meneghello2019,
alrawi2019} and smart-home automation risks~\cite{surbatovich2017}, but
not the LLM agent layer that now mediates between user intent and
physical actuation. MQTT-specific analyses~\cite{andy2017, harsha2018,
firdous2017, mishra2020} document authentication and access control gaps
in telemetry contexts; we show these gaps become safety-critical when
MQTT carries autonomous agent commands to physical actuators. Our testbed
uses WireGuard~\cite{donenfeld2017} via Tailscale, but our findings
confirm that transport-layer encryption does not address
application-layer coordination vulnerabilities (S1a--c).
\paragraph{Multi-agent coordination.}
Dorri et al.~\cite{dorri2017} applied blockchain to IoT multi-agent
trust, a solution ill-suited to resource-constrained edge
nodes where our NUC already exhibits $2.7{\times}$ latency overhead.
Calvaresi et al.~\cite{calvaresi2018} surveyed negotiation protocols
assuming static agent populations; LLM agent swarms add and remove nodes
dynamically. LLM-specific multi-agent work~\cite{talebirad2023, guo2024}
identifies coordination and trust as open problems but reasons about them
abstractly; we ground these concerns in empirical measurements from an
operational deployment controlling real IoT hardware.

\section{Conclusion}
Through empirical measurements on a three-node edge-local agent swarm,
we showed that deployment architecture, not model or prompt design, is
the primary determinant of security posture for agent-controlled IoT.
Edge-local MQTT eliminates structural data exfiltration and enables low actuation-to-audit delay, but provides zero cryptographic
provenance enforcement, exposes failover blind spots, and silently
degrades data sovereignty under stress. Two emergent failures
(coordination-state divergence and induced trust erosion) further
demonstrate that unauthenticated message buses are insufficient for
multi-agent coordination. Closing these gaps requires per-agent signing
keys, sovereignty-aware fallback policies, and network-level fast
roaming; we did not observe them in the frameworks used in our deployment or examined in this work. The systems
community should treat where agents run and how they communicate as
first-class security design choices, not implementation details
subordinate to model-level alignment.

%%
%% Acknowledgments
%%
\begin{acks}
This work was supported by the CHIST-ERA grant CHIST-ERA-22-SPiDDS-02 (GRAPHS4SEC). This work was conducted within the Networks and Systems Lab at Imperial College London. All adversarial experiments were conducted on the authors' own hardware in a closed local network with no connection to production systems or third-party infrastructure.

\end{acks}

%%
%% Bibliography
%%
\bibliographystyle{ACM-Reference-Format}
\bibliography{references}

%%
%% Appendix
%%
\appendix
\section{Extended Testbed and Measurements}
\label{sec:appendix}

Following the measurements reported in the main text (three-node testbed), the deployment expanded organically to five nodes. This appendix documents the extended topology and additional measurements that address scaling behaviour. 

The original three-node topology (Figure~\ref{fig:topology}) remains the basis for all main-text measurements. The two additional nodes described below joined the deployment subsequently.

\subsection{Extended Topology}

Two nodes were added to the original three-node swarm:

\begin{itemize}
  \item \textbf{Pixel 4} (Android 14, Google Pixel 4, Termux): mobile agent ``Darcy''
    (Grok 4.2). Darcy controls proprietary IoT applications that
    expose no external API, using UI automation through ADB and
    accessibility services. Darcy also accesses on-device sensors
    (camera, ambient light, gyroscope) that are physically unavailable
    to hub-resident agents, which is a key reason a single centralized agent
    cannot replace the distributed architecture. Darcy's role mirrors
    Percy's, enabling comparison across different models and hardware.
  \item \textbf{Intel NUC11} (x86, 4 cores, 7.5 GB RAM, Ubuntu 24.04): librarian agent ``Dewey''
    (observational only). Dewey maintains a Git repository as the
    canonical shared record of system coordination state. It subscribes
    to mirrored MQTT traffic and records externally relevant events
    with structured metadata (timestamp, sender, outcome). Events
    include task dispatches, device-state transitions, conflict
    resolutions, and recovery outcomes. Dewey does not issue commands
    and holds no device credentials; its role is purely observational.
    The repository provides three properties: (1)~versioned shared
    state that agents can inspect and recover after restart;
    (2)~attribution and history preserved in a diffable timeline; and
    (3)~stale-state detection via \texttt{base\_commit} comparison
    against the current HEAD.
\end{itemize}

\paragraph{Device adapters.}
Devices are not agents. Lights, speakers, cameras, and sensors are
thin, deterministic endpoints controlled by managers through
protocol-specific adapters. This keeps the agent count small and
avoids an ``agent per device'' design.

\paragraph{Dewey as operational response to S1b.}
The git-backed shared state plane maintained by Dewey directly
addresses the coordination-state divergence vulnerability (S1b)
identified in the main text (\S\ref{sec:provenance}). Rather than
relying on MQTT messages as the sole coordination medium (which
produces divergent local copies that drift silently) agents can
reference canonical state in the repository. Conflict detection
becomes explicit through Git's merge semantics rather than invisible.

\subsection{Scaling: Fan-out Latency}

To address the question of how the MQTT coordination layer scales
beyond three nodes, we measure fan-out latency as the number of
target agents increases from 1 to 4 (the full swarm excluding the
sender).

\begin{table}[h]
\centering
\caption{Fan-out latency: sequential publish to $k$ targets vs.\
broadcast ($N{=}50$ per configuration).}
\label{tab:fanout}
\begingroup
\setlength{\tabcolsep}{4pt}
\renewcommand{\arraystretch}{1.05}
\rowcolors{2}{gray!08}{white}
\begin{tabular}{lr}
    \toprule
    \rowcolor{white}
    Topology & Mean publish time \\
    \midrule
    1 target (original) & 35\,ms \\
    2 targets (fan-out)  & 41\,ms \\
    4 targets (full swarm) & 55\,ms \\
    Broadcast (single pub) & 35\,ms \\
    \bottomrule
\end{tabular}
\endgroup
\end{table}

Sequential fan-out grows linearly at ${\sim}5$\,ms per additional
target (Table~\ref{tab:fanout}), reflecting per-publish overhead
rather than broker contention. Broadcast (a single MQTT publish to a
shared topic) remains constant regardless of subscriber count, as the
broker handles fan-out internally. For safety-critical commands that
require all agents to observe the same action simultaneously,
broadcast is both faster and semantically stronger. The linear
sequential cost is modest for our five-node swarm but would become a
bottleneck at $N{>}20$ agents if per-agent addressing is required.

\end{document}